\documentclass{article}

\usepackage{PRIMEarxiv}
\usepackage{amsmath}
\usepackage[utf8]{inputenc} % allow utf-8 input
\usepackage[T1]{fontenc}    % use 8-bit T1 fonts
\usepackage{hyperref}       % hyperlinks
\usepackage{url}            % simple URL typesetting
\usepackage{booktabs}       % professional-quality tables
\usepackage{amsfonts}       % blackboard math symbols
\usepackage{nicefrac}       % compact symbols for 1/2, etc.
\usepackage{microtype}      % microtypography
\usepackage{lipsum}
\usepackage{fancyhdr}       % header
\usepackage{graphicx}       % graphics
\usepackage{booktabs}
\graphicspath{{media/}}     % organize your images and other figures under media/ folder

\title{Optimization of Tritium Breeding Ratio in a DT and DD Submersion Tokamak Fusion Reactor
}

\author{
  Vikram Goel \\
  United World College of South East Asia\\
  Singapore\\
  \texttt{vikramgoel12133@gmail.com} \\
   \And
  Soha Aslam \\
  Princeton University \\
  Princeton\\
  \texttt{sohhae@gmail.com} \\
  \And
  Sejal Dua \\
  Georgia Institute of Technology \\
  New York \\
  \texttt{sejaldua@gatech.edu}\\
}

\begin{document}
\maketitle

\begin{abstract}
The mass of stars is enough to confine a plasma to fuse light atoms, but this is not possible to engineer on Earth. Fortunately, nuclear engineering can rely on the magnetic confinement of a plasma using superconducting coils so long as the Tritium Breeding Ratio (TBR) is optimized. This paper will investigate some of the materials which can increase the rate at which Tritium is produced within the breeding blanket layer of Submersion Tokamak reactors, a design that uses magnetic confinement of a plasma in the shape of a torus to execute nuclear fusion. Using the Paramak Python module to model several geometries and OpenMC to run a simulation, it can be observed how neutron multipliers, enrichment, and the neutron energy spectrum affect TBR. This experiment will mainly observe different material choices that have been considered and their TBR based on their cross sections, dose rate, thermal properties and safety. By altering the neutron energy spectrum to account for DD and DT plasma, the difference in these compounds' Tritium breeding efficacy is noted. Neutron energy spectra are an important factor in optimising the TBR levels as the neutrons generated by the fusion reactions in the plasma interact with the breeder material in the blanket and produce tritium through the reaction with Lithium. Since Tritium is a rare isotope of hydrogen that is used as fuel in fusion reactions and has a short half-life, it is essential to produce tritium within the fusion reactor itself. Without the tritium breeding capability, it would not be feasible to generate energy via fusion. A TBR greater than unity indicates that the reactor can generate more tritium than it consumes, ensuring self-sufficiency in the tritium inventory. Since Tritium is the most reliable and efficient fuel for these reactors, optimising the TBR is of paramount importance in the long road to commercialization of nuclear fusion.
\end{abstract}

% keywords can be removed

\section{Introduction}
In the realm of energy production, nuclear fusion has emerged as a field of immense importance and potential as a promising source of clean and sustainable power. Fusion energy, simply put, is the energy released when light atomic nuclei combine to form heavier nuclei. This process occurs at incredibly high temperatures and pressures, replicating the conditions found in the core of stars. Unlike current energy sources, such as fossil fuels, it produces virtually no greenhouse gas emissions or long-lived radioactive isotopes. Furthermore, fusion energy has the potential to provide an almost limitless source of power, promising a sustainable and clean solution to our ever-growing energy needs.

To comprehend the various forms of nuclear fusion reactions, it is essential to explore the fundamental principles governing these processes. Fusion reactions can occur between different isotopes, with the most promising reactions involving deuterium (H2) and tritium (H3) nuclei. The fusion of these hydrogen isotopes releases a significant amount of energy and produces alpha particles as a byproduct. Additionally, alternative fusion reactions, such as the fusion of deuterium with helium-3 or boron-11, offer unique advantages and challenges, making them worthy of investigation, this paper will not delve into these though they are worthy of investigation.

To study and analyse the complex behaviour of nuclear fusion reactions, scientists often employ Monte Carlo simulations. These simulations utilise probabilistic methods to model the behaviour of particles within a system, providing insights into the dynamics of nuclear reactions. Monte Carlo simulations involve the generation of random numbers to determine the characteristics of each particle, allowing researchers to predict the behaviour and outcomes of fusion processes accurately. By simulating a large number of particles, scientists can obtain statistically significant results and optimise experimental designs, ultimately advancing our understanding of fusion energy and aiding in the development of practical fusion reactors.

\begin{equation}
    \label{eq:eq1}
    \text{Tritium Breeding Rate (TBR)} = \frac{\text{number of tritium atoms produced}}{\text{number of tritium atoms fused}}
\end{equation}

One of the most important statistics concerning fusion reactors is the Tritium Breeding Ratio, hitherto TBR. This is a representation of the ratio of the rate at which tritium is produced in a fusion reactor to the rate at which it is consumed \ref{eq:eq1}. Since Tritium is the most important fuel source in most reactors, it is essential to produce Tritium within the reactor itself: Tritium has a very short lifespan (12.3 years) and out of all the Hydrogen in the universe, tritium only makes up 10-18\% while deuterium sits at 0.2\%. This small quantity is meaningless and therefore tritium is made artificially using CANDU fission reactors so it is not viable to continuously add tritium to reactors, therefore, maintaining a TBR greater than unity is essential for commercialising DT nuclear fusion.
The main component of nuclear fusion reactors, specifically tokamaks, that is affiliated with the TBR is the breeding blanket. The breeding blanket is usually made of either Solids, Liquids (PbLi), Molten Salts (FLiBE), or pure Lithium. For this paper, a survey of two several breeder materials was undertaken as well as observing the impacts of Li6 enrichment and neutron multiplier concentrations.

\section{Methodology}
\label{sec:methodology}

\subsection{OpenMC, DAGMC, Paramak and Defining Computational Intensity}

Neutronics modelling and simulation is conducted using a monte carlo simulation tool, known as OpenMC. The OpenMC module simulates neutral particles (presently neutrons and photons) moving stochastically through an arbitrarily defined model that represents a real-world experimental setup \cite{rhhnfs15}. It was developed by the Computational Reactor Physics Group (CPRG) based at the Massachusetts Institute of Technology (MIT) and has been extensively utilised in the area of nuclear engineering.

For this research, a geometry for a Submersion Tokamak reactor was constructed using the Paramak Python module developed by Jonathan Shimwell \cite{sbdchstebmhmidppfs21}. This library makes use of parametric shapes, components and reactors, and is essential for constructing complex geometries that could be assigned different materials within the OpenMC model. Though OpenMC manipulates Boolean operators to construct simple geometries, it is difficult to construct a complex tokamak with a high degree of similarity to what is used in actuality through OpenMC alone. Upon constructing the geometry, Paramaks computer-aided design (CAD) would be converted into a ‘dagmc.h5m’ file, the format which works with OpenMC, using the \texttt{cad\_to\_dagmc} python module. Then, a neutronic simulation was conducted using a Python script. Within the OpenMC model, the different parametric shapes could be assigned to various elements and densities then, after introducing a neutron source a simulation using 10 rounds of 10,000 particles was conducted.

Within the model, interchanging between Lithium compounds was simple and separate tallies for these different blanket materials were generated using the 2.5 and 14 MeV neutron energy spectra.

\subsection{Monte Carlo Simulation}

OpenMC uses a method that resembles the MC21 monte carlo code (A neutron and photon transport code). To provide workability in a macro environment OpenMC uses an analogue estimator to obtain its tallies for reaction rates, using massive amounts of nuclear data, a survey is taken of the number of actual reactions that occur and then this is used for their estimates on what the reaction rates and therefore, the scores should be. 

This is represented by the OpenMC Analog Estimator Formula:

\begin{equation}
R_x = \frac{1}{W} \sum_{i \in A} w_i
\end{equation}

where Rx is the reaction rate for reaction x, i denotes an index for each event, A is the set of all events resulting in reaction x, and W is the total starting weight of the particles, and wi is the pre-collision weight of the particle as it enters the event \cite{rhhnfs15}. 

This analogue estimator is the simplest way to estimate and is easy to implement, using this OpenMC may use nuclear cross sections data for neutron multipliers and Li6 to generate its scores on the TBR, but this estimator has variance since low-probability reactions will require an immense amount of information and particles to calculate to a high degree of accuracy. Therefore, OpenMC also uses a collision estimator and a track length estimator to get around this issue.

The collision estimator does not simply add to a tally once a reaction has been carried out, it will use every collision to add to the score, for example, in the context of Tritium Breeding, it will not only add to the tally every time tritium is produced but every time there is a collision between the reactants that may or may not produce tritium.

\section{Materials}

\begin{figure}[hbt!]
  \centering
  \includegraphics[width=0.8\linewidth]{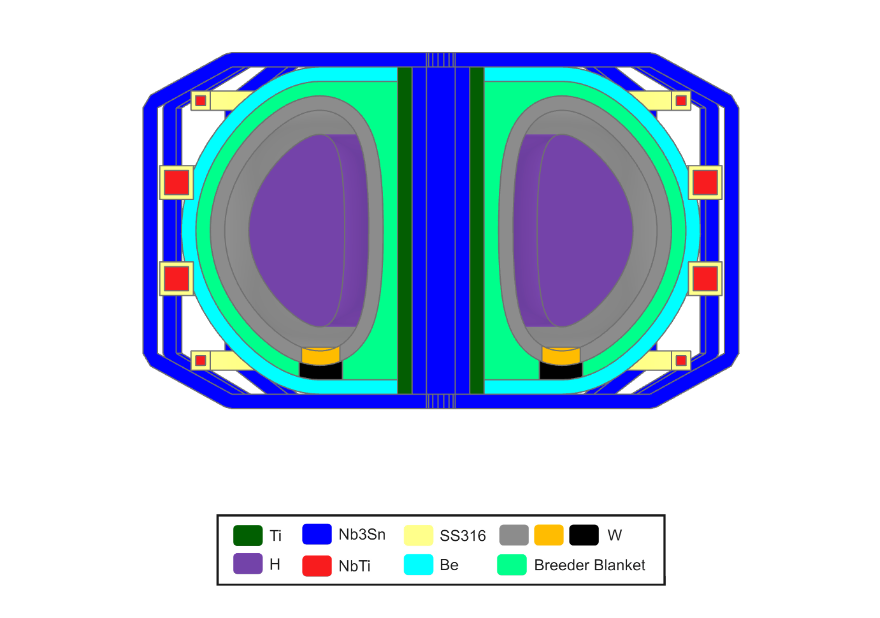}
  \caption{Diagram of the Geometry of the Submersion Tokamak}
  \label{fig:fig1}
\end{figure}

The design used for this paper is the Tokamak, it is the leading style of fusion reactor and its geometry is shown in Figure \ref{fig:fig1}. The magnetic coils that were created for this simulation were Niobium Titanium (NbTi) coils chosen due to their superconductivity, high critical current density, fabrication availability and mechanical strength. These attributes mean that it has zero electrical resistance, allows a high current to pass through without reverting to its conducting state, can be easily made into wires and coils and importantly, can withstand the immense electromagnetic forces within a tokamak. We also used a similar metal, Nb3Sn for the outer toroidal field coils for the same reasons, Tungsten for the first wall and components directly adjacent to the plasma due to its thermal properties and Titanium supports. A Beryllium blanket wall was used, to maximise neutron multiplication and Stainless Steel 316 for the poloidal field coils casing as it tolerates elevated temperatures.

For this simulation, we observe how these different materials compare when used as breeder blankets in producing Tritium. This will be done with both Deuterium-Tritium (D-T) and Deuterium-Deuterium (D-D) fuel. While D-T is a lot more powerful and yields more energy and tritium, the deuterium isotope of hydrogen is far more abundant. It has a longer life making it less expensive and more feasible for implementation. The two hydrogen plasmas will react to form 14.1 MeV and 2.5 MeV neutrons as shown:

\begin{equation}
D + T \rightarrow [\alpha + 3.5 MeV] + [n + 14.1 MeV]
\end{equation}
\begin{equation}
D + D \rightarrow 3He + [n + 2.5 MeV]
\end{equation}

A Beryllium (Be) or Lead (Pb) neutron multiplier will increase the number of neutrons by initialising the (n,2n) reaction. For the breeder blanket materials, a variety of Lithium-based alloys and compounds were used, Lithium is especially important for the breeder blanket as it reacts with neutrons to produce tritium and, when paired with a Beryllium (Be) neutron multiplier, it will efficiently produce tritium at a rate higher than unity (TBR = 1). It is the most reliable element to use in breeder blankets as its only two stable naturally occurring isotopes react with the neutrons as shown:

\begin{equation}
% TODO: cite reference 
^6Li + n \rightarrow \alpha + T + 4.78 MeV
\end{equation}
\begin{equation}
^7Li + n \rightarrow \alpha + T + n + 2.47 MeV
\end{equation}

Energy Production and extraction within these tokamak reactors is a tedious engineering task: The kinetic energy of the incident neutrons must be converted into thermal energy for it to be excavated by the coolant. The breeder blanket can augment the incident energy via exothermic reactions including the $^6Li(n,t)^4He$ reaction which has a Q value of 4.79MeV. 

\begin{figure}[hbt!]
  \centering
  \includegraphics[width=0.8\linewidth]{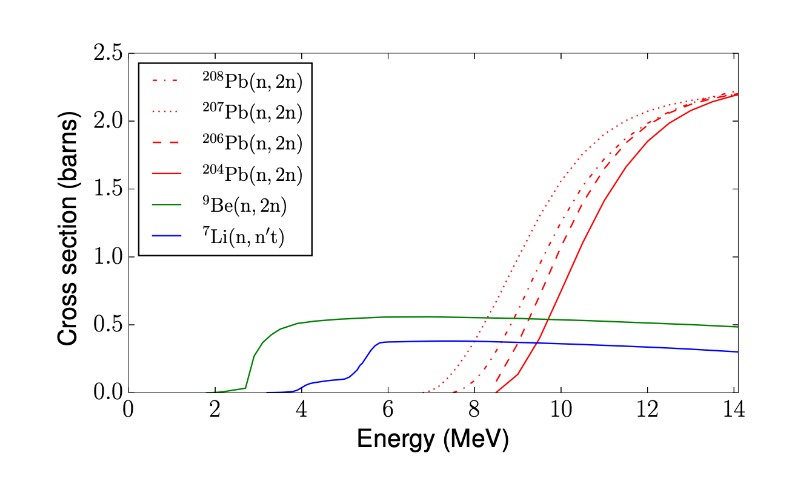}
  \caption{Cross Section vs Energy \cite{chadwick}}
  \label{fig:fig2}
\end{figure}

Assuming the use of Pb multipliers, DT fusion devices will perform better multiplication due to high neutron multiplication cross-section at those energies as shown in Figure \ref{fig:fig2}.

\section{Results}

Table \ref{tab:table} of the Tritium Breeding ratio for several Lithium-Based materials that are used as a breeder blanket in a Submersion Tokamak with both 14.1 MeV and 2.5 MeV neutrons, these were obtained upon the completion of the monte carlo simulation:

% \usepackage{booktabs}
\begin{table}[]
\centering
\caption{TBR Results}
\label{tab:table}
\begin{tabular}{@{}lrr@{}}
\toprule
\textbf{Lithium-Based material}          & \textbf{\begin{tabular}[c]{@{}r@{}}Tritium Breeding Ratio \\ Tally (14MeV)\end{tabular}} & \textbf{\begin{tabular}[c]{@{}r@{}}Tritium Breeding Ratio \\ Tally (2.5MeV)\end{tabular}} \\ \midrule
\textbf{Li2ZrO3 (Lithium Zirconate)}     & 0.974073                                                                                 & 0.490369                                                                                  \\ \midrule
\textbf{Li2TiO3 (Lithium Titanate)}      & 0.980266                                                                                 & 0.501079                                                                                  \\ \midrule
\textbf{Li4SiO4 (Lithium Orthosilicate)} & 1.00325                                                                                  & 0.525374                                                                                  \\ \midrule
\textbf{LiCl (Lithium Chloride)}         & 1.04656                                                                                  & 0.539386                                                                                  \\ \midrule
\textbf{FLiBe (LiF and BeF2)}            & 1.08196                                                                                  & 0.569398                                                                                  \\ \midrule
\textbf{Li2O (Lithium Oxide)}            & 1.09383                                                                                  & 0.542774                                                                                  \\ \midrule
\textbf{Li17Pb83 (90\% Li6)}             & 1.14519                                                                                  & 0.570423                                                                                  \\ \midrule
\textbf{Li (Pure Lithium)}               & 1.10905                                                                                  & 0.543946                                                                                  \\ \bottomrule
\end{tabular}
\end{table}
\subsection{Lithium-6 Concentration}

It appears as though the concentration of Lithium-6 is the factor that most affects the TBR as it has a high cross-section for neutron absorption and is the isotope of Lithium that produces the Tritium via Li6(n, alpha)H3, so the TBR of breeder material Li17Pb83 was measured over different levels of enrichment of Lithium-6:

\begin{figure}[hbt!]
  \centering
  \includegraphics[width=0.8\linewidth]{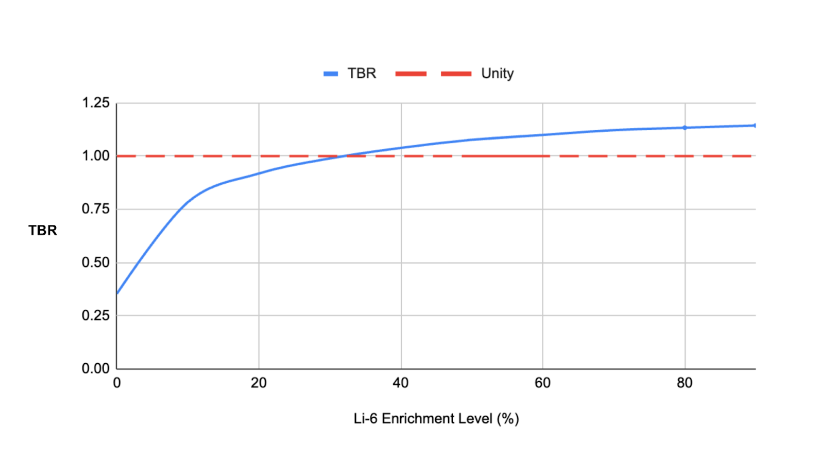}
  \caption{Enrichment vs TBR}
  \label{fig:fig3}
\end{figure}

Li17Pb83 is usually used as a breeder with a Li-6 concentration of 90\% or higher, this is feasible as it has a low overall Lithium concentration and high reliance on the lead as a multiplier. Figure \ref{fig:fig3}, depicting the relationship between Li-6 enrichment level and TBR, can be described by this equation with under a 5\% margin of error for the largest outlier:

\begin{equation}
Y = 0.34 \log{[1.4(X)  +  1.11]} \quad \quad , {0<X<1}
\end{equation}

where Y is the TBR and X is the Li6 enrichment as a decimal.

Since lead, as shown in Figure \ref{fig:fig2}, has a high cross section for neutron multiplication by the (n,2n) reaction at ~6.5 MeV or higher, D-D neutrons (<6.5MeV) perform worse at a Tritium breeding rate of less than half of what it is for D-T neutrons (>6.5MeV). 
The answer considering the TBR is clear: The higher the concentration of Lithium and thereby, Lithium-6, the greater the raw Tritium Breeding, but this is not to say that pure Lithium-6 is the best material to use as a breeder - there are many other factors to consider, Pure Lithium has inferior thermal properties when compared to Lithium Titanate or Lithium Orthosilicate and is unable to withstand immense heat, it is also more unstable than these compounds and therefore poses a safety risk when used in the breeder blankets. It is also not as strong mechanically and is far more pliable in its pure state, a reason why alloys or compounds prove beneficial. The material must withstand extreme temperatures, and electromagnetic forces so Lithium is simply not as reliable. 
Furthermore, as this graph for Li-6 concentration vs TBR is logarithmic, there is little need, in most cases to enrich further than 40\%, increasing Li-6 concentration to this level changes the TBR by about 0.7, but a further increase to 90\% will only change the TBR by about 0.1. 

\subsection{Ceramic, Liquid and Molten Salt Blankets}

The breeder materials observed using the OpenMC simulation are ceramic liquid and molten salt breeders. Li4SiO4 (Lithium Orthosilicate), Li2O (Lithium Oxide), Li2ZrO3 (Lithium Zirconate), and Li2TiO3 (Lithium Titanate) make up the ceramic breeders, they all have TBRs close to 1, ranging from 0.97 to 1.0 (nearest hundredth) and their energy multiplication are also similar with Lithium Titanate coming out on top at 1.15 and the lowest at 1.05 \cite{SAWAN20061131}. They have similar thermal properties including their melting point and thermal expansion. Their thermal properties and resistance to strong electromagnetic forces within the tokamak make them especially useful over pure Lithium or weaker alternatives. Lithium vacancy in Li4SiO4 may interact with tritium resulting in some confinement, an essential outcome for extracting energy and recycling Tritium efficiently. The formation energies of the vacancy-tritium complexes are in the range of 0.41 - 1.28 eV under oxygen-rich conditions \cite{slgxzygy15}. 

Liquid breeder materials have several qualities that make them superior when compared to their ceramic counterparts including thermal conductivity, fabrication availability and the ease at which they can be removed for Tritium Extraction \cite{tmrs88}. PbLi100, a liquid breeder that we simulated in addition to those in the table above, has a TBR of 1.24918 with D-T and 0.639227 with D-D, topping both categories.

\begin{table}[hbt!]
\centering
\caption{PbLi100 TBR}
\label{tab:table2}
\begin{tabular}{@{}lll@{}}
\toprule
PbLi100 & 1.24918 (DT TBR) & 0.639227 (DD TBR) \\ \bottomrule
\end{tabular}
\end{table}

Though Lead Lithium with its high TBRs may seem ideal, they suffer from high inefficiency as the next chapter describes.

\subsection{Dose Rates}

Even Li17Pb83 and other PbLi-based breeders have the highest TBRs at around 1.2 in our simulation. But still, these liquid breeders with high cross sections for neutron multiplication and tritium production may not be as efficient as FLiBe, the leading molten salt blanket. FLiBe is usually a mixture of LiF and BeF2 and is one of the molten salt blankets. Though it may not have such a high TBR, it is incredibly efficient in absorbing radiation. Below is a graph illustrating how FLiBe and PbLi compare with Lithium and other blankets in terms of their Dose Rates over time:

\begin{figure}[hbt!]
  \centering
  \includegraphics[width=0.6\linewidth]{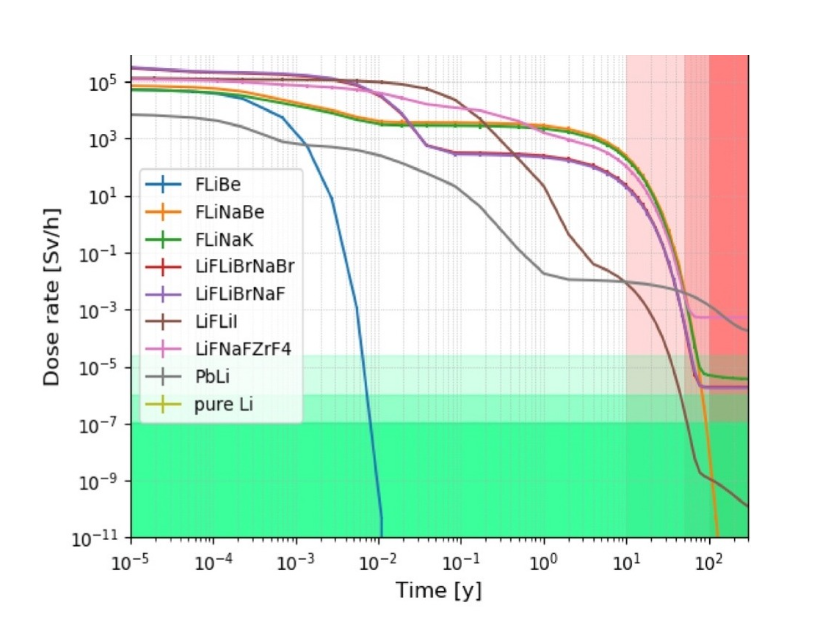}
  \caption{Neutronic comparison of liquid breeders for ARC-like reactor blankets \cite{SEGANTIN2020112013}}
  \label{fig:fig4}
\end{figure}

The green areas in Figure \ref{fig:fig4} represent the recycling limits, and only FLiBe and a few other salts can meet these limits in a short amount of time whereas PbLi is far more inefficient. FLiBe has a high heat capacity and a high melting point, attributes that may contribute to its efficiency \cite{SEGANTIN2020112013}.

\subsection{Neutron Multiplier Concentration}

Another question regarding the materials to use in a breeder blanket to maximise tritium breeding and efficiency is the ratio of neutron multipliers to the Lithium material, the two materials needed for the two reactions used to produce the hydrogen isotope.

This chart illustrates the TBR of a Lithium based blanket paired with a simple Lead neutron multiplier over different ratios of Pb to Li. (This used a fixed mass of Li6, therefore, the Li6 was always 0.2 x total breeder blanket mass, it was adjustable in the simulation by calculating the required fraction out of total Lithium to make it 20\%, for example, 90\% Lithium x 22.2..\% = 20\% Lithium-6):

\begin{figure}[hbt!]
  \centering
  \includegraphics[width=0.8\linewidth]{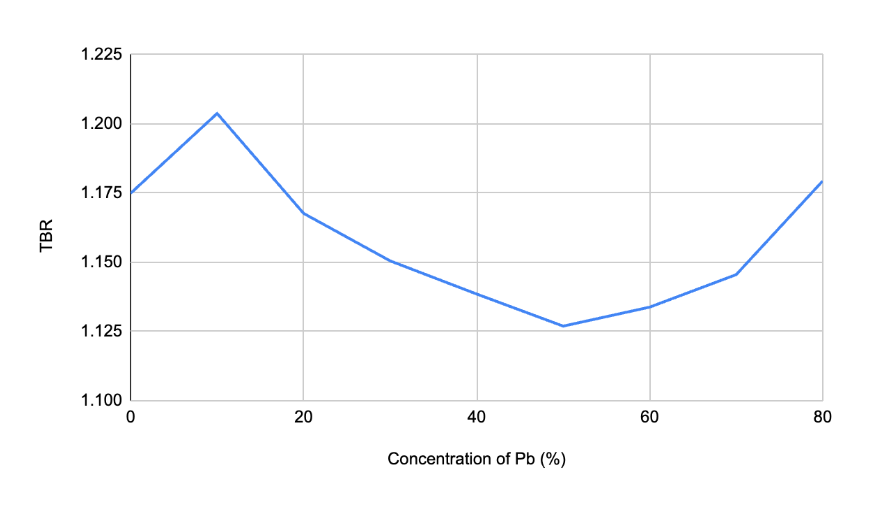}
  \caption{TBR against Lead concentration in Breeder Blanket}
  \label{fig:fig5}
\end{figure}

The parabolic shape of these results is due to there being a constant mass of Lithium-6, this means that at lower lead concentration, there is no extra Li-6 that could significantly increase the TBR and at higher lead concentration, this helps the TBR to increase as there is a better concentration of multipliers. In Li17Pb83, there is a much higher concentration of lead, but this is compensated with high Li6 enrichment, therefore, it makes sense to assume a relatively constant amount of Lithium 6 should be present.

\section{Conclusion}

This paper has observed the effects, advantages and results of using different types of blanket materials, different levels of Li-6 enrichment and concentration of neutron multipliers. We have also addressed other concerns such as thermal properties, energy multiplication and dose rates/efficiency.

In the matter of the TBR, the best blanket material appears to be Li17Pb83, it is the breeder that, unlike FLiBe, has no toxicity \cite{rh17} and unlike PbLi100, is not so inefficient in terms of dose rate. It also has relatively low concentrations of Lithium, a scarce material that may prove difficult to obtain in the coming decades. This liquid breeder also comes with the advantages that liquid breeders have over their solid counterparts as previously discussed and it has a very high TBR when compared to most other materials. FLiBe was a very successful breeder especially when one takes into account the dose rates and apart from the hazard risks with this breeder it may be the most successful one that was used.

To further advance our understanding of the topic of breeder blankets and tritium breeding, further research into other attributes must be explored such as different thicknesses of breeders and alternate fuel sources such as boron-11 or helium 3. It is also essential to address the issues of tritium extraction and confinement aside from uranium beds that may be proliferation concerns. Construction of several breeds of tokamaks and even stellarator reactors using Paramak or the new Parametric Stellarator python libraries to pair with an OpenMC simulation could also be researched to further understand the TBRs of different blanket materials in more than one type of reactor. 

The field of computational application for nuclear engineering should be used to predict and theorise on how these reactors and particles interact and will, in the future, be of great use as fusion becomes the energy source for our planet’s fossil fuel-free future.

\section*{Acknowledgments}
This was supported by my mentor, Ms. Soha Aslam who instructed and guided my research process, teaching me about how to use the python packages I needed and about this field as a whole.

This was also made possible by the Paramak library, developed by Dr. Jonathan Shimwell.

%Bibliography
\bibliographystyle{unsrt}  
\bibliography{references}

\end{document}